# Digital defocus aberration interference for automated optical microscopy

Haowen Zhou[1,†,*], Shi Zhao[1,†], Yujie Fan[2], Zhenyu Dong[1], Oumeng Zhang[1], Viviana Gradinaru[2], and Changhuei Yang[1]

[1]Department of Electrical Engineering, California Institute of Technology, Pasadena, CA, USA
[2]Division of Biology and Biological Engineering, California Institute of Technology, Pasadena, CA, USA
[†]These authors contributed equally to this work.
[*]hzhou7@caltech.edu

## Abstract

Automation in optical microscopy is critical for enabling high-throughput imaging across a wide range of biomedical applications. Among the essential components of automated systems, robust autofocusing plays a pivotal role in maintaining image quality for both single-plane and volumetric imaging. However, conventional autofocusing methods often struggle with implementation complexity, limited generalizability across sample types, incompatibility with thick specimens, and slow feedback. We observed that the digitally summed Fourier spectrum of two images acquired from two-angle illumination exhibits interference-like fringe modulation when the sample is defocused. These digital fringes correlate directly with defocus through a physics-based relation. Based on this principle, we developed an automatic, efficient, and generalizable defocus detection method termed digital defocus aberration interference (DAbI). Implemented with a simple two-LED setup, DAbI can quantify the defocus distance over a range of 443 times the depth-of-field (DoF) for thin samples and 296 times for thick specimens. It can additionally extend the natural DoF of the imaging system by 20 folds when integrated with complex-field imaging. We demonstrated the versatile applications of DAbI on brightfield, complex-field, refractive index, confocal, and widefield fluorescence imaging, establishing it as a promising solution for automated, high-throughput optical microscopy.

## Introduction

Optical microscopy has become an indispensable tool in life and medical sciences for visualizing diverse structural, chemical, and biological information across various scales[1]. Despite its widespread utility, conventional optical microscopy often involves labor-intensive manual operations, such as focusing adjustments, exposure settings, and sample positioning[2]. These manual processes are time-consuming and inconvenient, significantly limiting the scalability of microscopy, particularly for high-throughput applications. Consequently, there is a strong need to automate these operations, especially focus control for two-dimensional (2D) and three-dimensional (3D) samples, as precise focusing directly influences image quality and critically impacts the accuracy of subsequent biological and medical analyses.

The automation of focusing or autofocusing can be accomplished by a number of different techniques for 2D thin samples. We define autofocusing as the task of determining the defocus distance and actuating the sample (or microscope) to place it in the correct focal plane (Fig. 1a). In general, the defocus distance determination is the primary challenge in autofocusing as the involved mechanical actuation is straightforward. The different autofocusing methods for defocus distance determination can be classified into several primary categories. The first category consists of deep learning methods[3–7]. These methods have recently been actively applied in 2D autofocusing to estimate defocus distance from several or even a single image. Nevertheless, these methods have limited search range (e.g. 17× depth-of-field (DoF) for 20×/0.75NA objective[4]). They also heavily rely on training datasets and suffer from generalization issues. The second category consists of passive image-based autofocusing techniques[8–11]. These methods calculate a focusing metric distribution from multiple images taken at different axial positions to determine the correct focal plane. These methods necessitate multiple image acquisitions within a pre-defined range and will fail if the correct focal plane falls outside the range. Moreover, these methods lack a universal metric for diverse samples[12]. As such, these methods are inefficient and non-robust. The last category consists of active autofocusing techniques, such as tilted sensor[13,14], triangulation[15,16], interferometry[17,18] , and phase detection[19,20]. These methods employ auxiliary optical or mechanical hardware components to detect the defocus distances. These methods involve complex hardware configurations and often struggle to generalize across different optical imaging modalities.

Of these methods, the dual illumination method[21–24], which uses two-opposite-angle illumination to find defocus distance from image lateral displacement, stands out for its simplicity, speed and generalizability. Yet even this method has its own limitations. At small defocus distances (<46× DoF for 20×/0.75NA objective), this method fails when the two-copy of the image cannot be separated apart to retrieve defocus distance[22]. Additionally, this method suffers from poor accuracy at large defocus detection due to diffraction effect and image deformation[24], limiting its broad applications.

Unsurprisingly, autofocus automation for optical microscopy imaging of 3D samples is an even more challenging task. Here, autofocus is aimed at identifying the nominal central plane of the 3D sample for automated volumetric imaging. By its nature, a 3D sample would intrinsically contain multiple planes of interest, and the contributions from these planes would obfuscate the process. To date, 3D autofocusing has only been demonstrated by using phase detection method[25], with limited autofocusing range (70× DoF, with 10×/0.60NA objective) and complex system designs. In most situations, focus tuning for 3D samples still rely on manual adjustments and visual inspections.

In this paper, we report the interference-like fringes arising from the digital sum of the Fourier spectra of two intensity images acquired from closely spaced oblique illumination angles from an out-of-focus sample. We analytically model the formation of these fringes from wave optics and optical aberration theory. The resulting fringe patterns correlate directly with sample defocus. Using this relation, we develop a robust and automated defocus estimation method that integrates tailored illumination design, a virtual defocus aberration strategy, and fringe-curvature compensation. We term this approach digital defocus aberration interference (DAbI) method. Compared with the dual-LED method, DAbI enables accurate defocus estimation across both small and large defocus distances and provides reliable autofocusing for thick samples.

DAbI enables autofocusing for both 2D and 3D samples by simply integrating two LEDs into microscopes. With DAbI, we achieved significant autofocusing capabilities, with ~1800 μm autofocusing range (443× DoF, with 20×/0.40NA objective) for 2D thin samples. Its use for thick samples autofocusing is even more significant. It offers a robust and efficient way to determine a 3D object's central plane, providing an autofocusing range of ~1200 μm (296× DoF, with 20×/0.40NA objective) for thick samples (40-150 μm in thickness), covering weakly/moderate scattering samples. DAbI also offers high efficiency and low photon toxicity, requiring the capture of only two images at a single plane with micro-watt level illuminations in our demonstration experiments. Its simplicity, utilizing just two LEDs, facilitates easy integration into customized or commercial imaging systems. Furthermore, the physics-based foundation of DAbI ensures high fidelity and broad applicability across diverse biomedical samples. We demonstrated the application of DAbI across a broad range of optical imaging modalities, including widefield fluorescence microscopy, quantitative phase imaging, brightfield transmission microscopy (BFM), refractive index tomography, and confocal microscopy.

The use of DAbI extends beyond autofocusing, and we have also demonstrated that DAbI can be applied to digital refocusing. The overall goal of digital refocusing is similar to autofocusing in that it seeks to return an in-focus image[26,27]. Digital refocusing differs from autofocusing in that it eliminates the mechanical actuation step altogether and instead uses the raw image data to numerically reconstruct the in-focus image (Fig. 1b). Digital refocusing can also be interpreted as the computational extension of a system's natural DoF, and its performance is benchmarked by the effective multiplicative extension factor by which the system's natural DoF has been increased. Digital refocusing is generally accomplished by the joint design of optics and algorithms, such as point spread function engineering[28–31] and illumination coding[32–34]. Despite widespread adoption, digital refocusing systems typically demand precise system alignment and offer only moderate extension of the effective DoF. Deep learning methods, particularly generative models[35–37], have been explored to computationally refocus defocused images by sampling distributions learned through neural networks[38]. However, these methods often face challenges related to fidelity and generalization, particularly when applied to samples outside the trained distribution.

Here, we show that DAbI can integrate seamlessly with existing complex-field imaging techniques, substantially extending the DoF of conventional system without additional data collection. In our experiments, we digitally extended the effective DoF of complex-field imaging by 20-fold compared to conventional systems, and a 4.8-fold improvement over the state-of-the-art method[39].

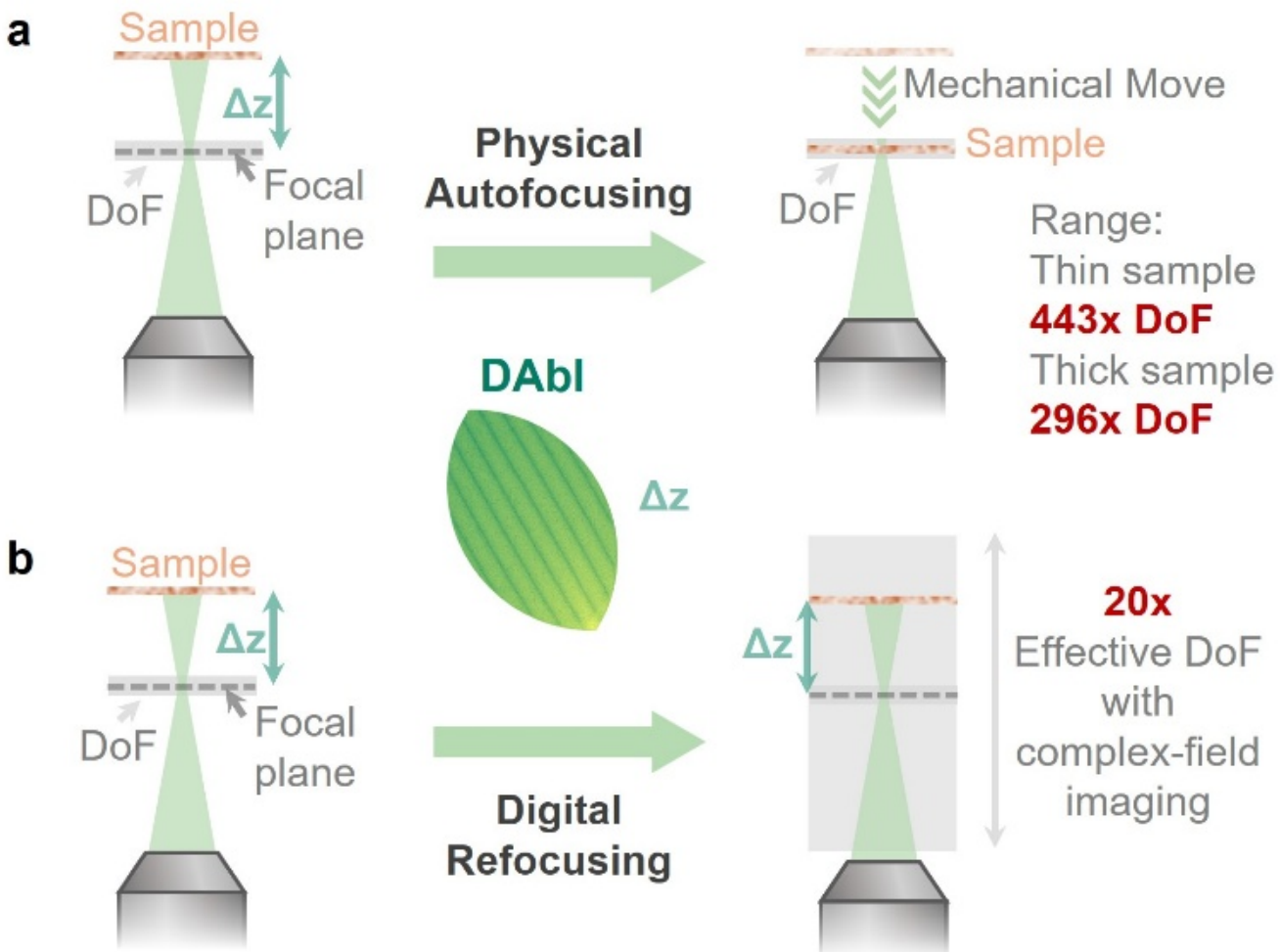

**Fig.1 | Illustration of autofocusing and digital refocusing in optical microscopy with DAbI.** The defocus Δz in optical microscopy is when the sample content lies outside the axial range of the system's depth-of-field (DoF). **a,** DAbI assists the automation of optical microcopy by determining the defocus distance Δz and actuating the sample to place it in the correct focal plane. In our demonstrated experiments, DAbI achieved 443 times DoF autofocusing range for 2D thin samples and 296 times DoF for 3D thick samples. **b,** When integrated with complex-field imaging, DAbI can also provide the defocus value as a prior to help digitally reconstruct high-quality in-focus images. With DAbI's assistance, the natural DoF of the imaging system can be extended 20 times.

## Results

### Principle of DAbI

DAbI retrieves defocus values using two-LED transmissive illuminations (Fig. 2). Specifically, two LEDs sequentially illuminate the sample from two different angles with certain separation. These two angles should be no larger than the maximum acceptance angle of the imaging system, defined by its numerical aperture (NA). The separation between two angles should be maintained at certain range (See Supplementary Note 3). These two oblique illuminations, $e^{-j\mathbf{k}_{i\perp}\mathbf{r}_{\perp}}$ (where $\mathbf{k}_{i\perp}, i = 1,2$, represent the illumination wavevectors and $\mathbf{r}_{\perp} = (x, y)$ denotes spatial coordinates), interact with sample and result in two intensity images through the imaging system.

We first analyze the scenario for 2D thin samples (Details in Supplementary Note 1). The Fourier transform of each intensity measurement comprises four parts: (1) the unscattered light contribution represented as a delta function; (2) the convolution of the shifted delta function with the scattered spectrum; (3) the complex conjugate of the second term with frequency shifts and inverses; and (4) the autocorrelation of the scattered spectrum. The four terms can be expressed in Eq. 1:

$$\mathcal{F}\{I_i\} = \delta(\mathbf{k}_\perp) + \hat{\Psi}(\mathbf{k}_\perp) \otimes \delta(\mathbf{k}_\perp + \mathbf{k}_{i\perp}) + \hat{\Psi}^*(-\mathbf{k}_\perp) \otimes \delta(\mathbf{k}_\perp - \mathbf{k}_{i\perp}) + \hat{\Psi}(\mathbf{k}_\perp) \otimes \hat{\Psi}^*(-\mathbf{k}_\perp),\ i = 1,2 \tag{1}$$

where $I_i$ is the $i^{th}$ intensity measurement; $\hat{\Psi}(\mathbf{k}_\perp) = \hat{O}(\mathbf{k}_\perp - \mathbf{k}_{i\perp})\hat{P}(\mathbf{k}_\perp)$ is an auxiliary function defined in terms of spatial frequency coordinates $\mathbf{k}_\perp$, with $\hat{O}(\mathbf{k}_\perp - \mathbf{k}_{i\perp})$ as the scattered sample spectrum shifted by $\mathbf{k}_{i\perp}$, and $\hat{P}(\mathbf{k}_\perp)$ as the coherent transfer function; $\delta(\mathbf{k}_\perp)$ is the Dirac delta fuction; $\mathcal{F}\{\cdot\}$ is Fourier transform operation; and $\otimes$ indicates convolution. Notably, the second term in Eq. 1, $\hat{\Psi}(\mathbf{k}_\perp) \otimes \delta(\mathbf{k}_\perp + \mathbf{k}_{i\perp}) = \hat{O}(\mathbf{k}_\perp)\hat{P}(\mathbf{k}_\perp + \mathbf{k}_{i\perp})$, can be interpreted as the scattered sample spectrum modulated by a shifted coherent transfer function (Supplementary Note 1.2).

Subsequently, if we sum up the Fourier transform of two intensity measurements $\mathcal{F}\{I_1\} + \mathcal{F}\{I_2\}$, the resulting spectrum constitutes a linear combination of the individual terms described in Eq. 1 (Fig. 2a). Due to the constraints of $\hat{P}(\mathbf{k}_\perp)$, these terms occupy distinct regions with finite support. Therefore, there exists a region that only consists of the second and fourth term in Eq. 1 from two illumination angles, designated as the "overlap region" (highlighted in white-shaded region of Fig. 2a,b). In this overlap region, if the sample is not highly absorptive or scattering, the fourth terms ($i = 1,2$) in Eq. 1, corresponding to the autocorrelation of the scattered spectrum, are negligible (Supplementary Note 1). In contrast, the second terms ($i = 1,2$) in Eq. 1 are dominant. The interplay of these two second terms introduce interference-like fringes. Thus, the spectra amplitude in the overlap region can be mathematically approximated as (Fig. 2b):

$$|\mathcal{F}\{I_1\} + \mathcal{F}\{I_2\}| = \left|\hat{A}(\mathbf{k}_\perp)e^{j\hat{\eta}(\mathbf{k}_\perp)}\right|\left|e^{j\hat{\phi}(\mathbf{k}_\perp+\mathbf{k}_{1\perp})} + e^{j\hat{\phi}(\mathbf{k}_\perp+\mathbf{k}_{2\perp})}\right| \propto \sqrt{2 + 2\cos\left(\hat{\phi}(\mathbf{k}_\perp + \mathbf{k}_{1\perp}) - \hat{\phi}(\mathbf{k}_\perp + \mathbf{k}_{2\perp})\right)} \tag{2}$$

where $\hat{A}(\mathbf{k}_\perp)e^{j\hat{\eta}(\mathbf{k}_\perp)} = \hat{O}(\mathbf{k}_\perp)$ is the complex-valued sample spectrum with amplitude $\hat{A}(\mathbf{k}_\perp)$ and phase $\hat{\eta}(\mathbf{k}_\perp)$; $\hat{\phi}(\mathbf{k}_\perp)$ represents the phase of coherent transfer function containing the defocus aberration. Eq. 2 illustrates the formation of fringes from the interference of two shifted defocus aberration at the pupil plane. The absolute defocus distance $\Delta z$ can be accurately determined by identifying the valleys (local minima) within the cosine fringe patterns:

$$|\Delta z| = \frac{|n\pi|}{\left|\sqrt{k_0^2 - (\mathbf{k}_{\mathbf{v}\perp} + \mathbf{k}_{1\perp})^2} - \sqrt{k_0^2 - (\mathbf{k}_{\mathbf{v}\perp} + \mathbf{k}_{2\perp})^2}\right|} \tag{3}$$

where $k_0 = \frac{2\pi}{\lambda}$ is the wave number corresponding to wavelength $\lambda$; $n = 2m - 1, m \in \mathbb{Z}$ specifies the $n^{th}$ order fringe valleys; and $\mathbf{k}_{\mathbf{v}\perp}$ satisfies $\hat{\phi}(\mathbf{k}_{\mathbf{v}\perp} + \mathbf{k}_{1\perp}) - \hat{\phi}(\mathbf{k}_{\mathbf{v}\perp} + \mathbf{k}_{2\perp}) = n\pi$.

When the defocus distance $\Delta z$ is small, the resulting fringes become sparse and may even vanish. To address this limitation, we introduce a virtual defocus aberration strategy that digitally reformulates the fringe patterns (Supplementary Note 1.6, 4). This strategy additionally enables unambiguous determination of the defocus direction over both small and large defocus ranges (Supplementary Note 1.6, 4).

In Eq. 3, the term $\sqrt{k_0^2 - (\mathbf{k}_{\mathbf{v}\perp} + \mathbf{k}_{1\perp})^2} - \sqrt{k_0^2 - (\mathbf{k}_{\mathbf{v}\perp} + \mathbf{k}_{2\perp})^2}$ introduces curved fringe patterns due to its intrinsic nonlinearity, especially pronounced in high NA settings (Supplementary Note 1.5). This nonlinearity can lead to errors in defocus estimation. To tackle this challenge, we designed a fringe curvature compensation strategy – an iterative procedure in our defocus estimation algorithm – to compensate for this nonlinearity (Supplementary Note 1.5, 4). Furthermore, the validity of Eq. 3 relies on the defocus being the dominant aberration. For scenarios where other aberrations, such as spherical or astigmatic aberrations, are strong, a one-time system aberration pre-calibration is required. A comprehensive introduction to the defocus estimation algorithm is presented in Supplementary Note 4 with explanations and pseudocode.

The retrieved defocus distance from DAbI can also serve as a defocus prior to perform digital refocusing with various complex-field imaging techniques[27,39,40]. This defocus prior generates a Zernike defocus aberration in the phase of the pupil function $\hat{P}(\mathbf{k}_\perp)$. This updated pupil function is then embedded into the reconstruction algorithm for computing high-quality images[41].

For 3D thick samples, according to the Fourier diffraction theorem[42], the second terms in Eq. 1 are distributed over 3D spherical caps (Supplementary Note 2). Thus, the overlap region transitions from a broad area in 2D cases to a narrow line (Dashed orange line in Fig. 2a) for 3D cases (Supplementary Note 2). Given the finite sample thickness[43], the two sample spectra from the two intensity measurements can be considered mostly identical but with a small difference around this narrow overlap line. Such differences degrade the fringe visibility, becoming more pronounced with increasing distance from this line, thereby reducing fringe clarity. Clear interference-like fringes, therefore, only appear in proximity to this narrow line, where Eq. 3 remains valid. For thicker or more scattered samples, the region displaying clear fringes become narrower, constraining the effective application of DAbI. This phenomenon parallels the coherence area concept in interferometry, where fringe visibility gradually diminishes due to unstable phase differences.

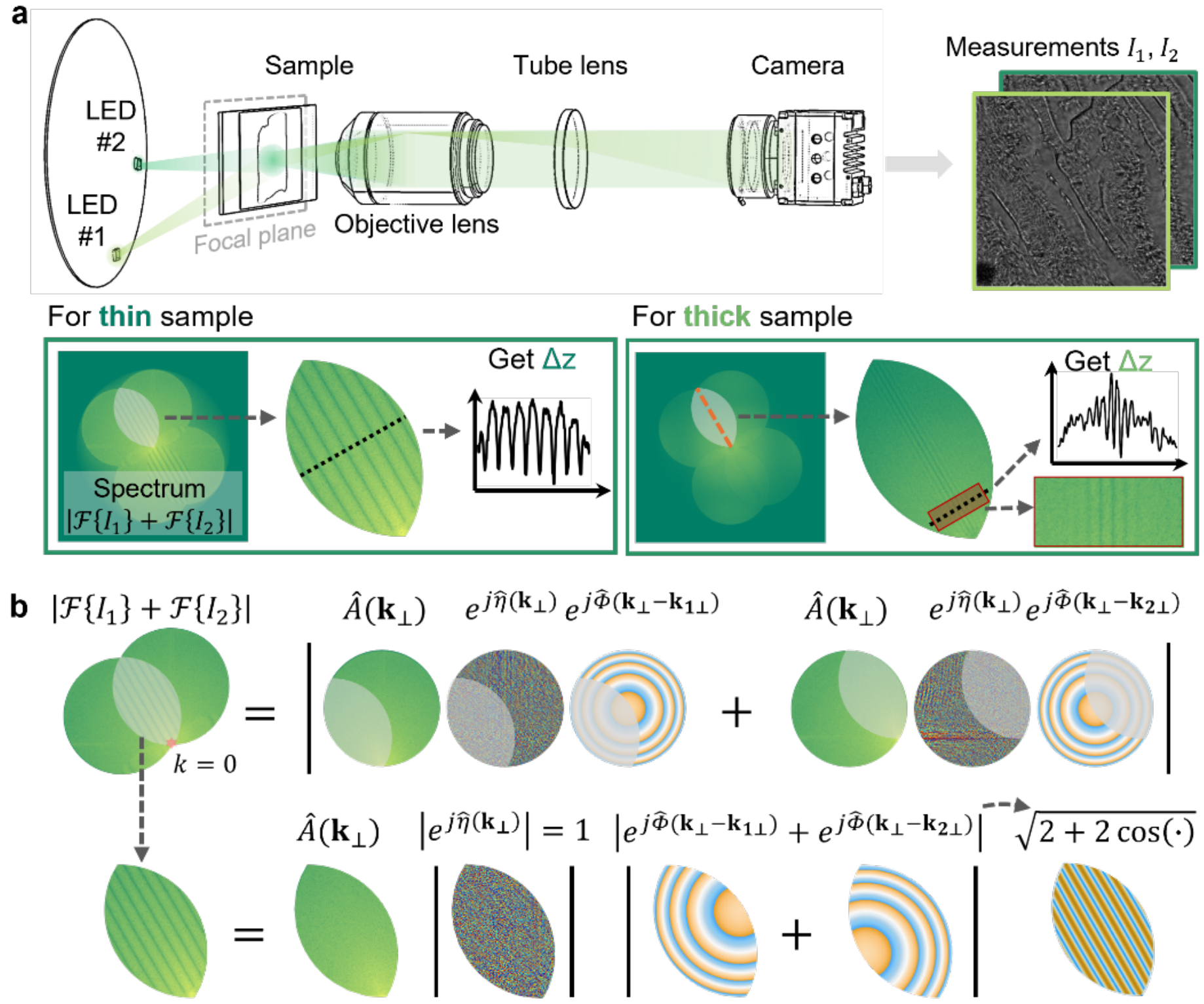

**Fig.2 | Principle of DAbI. a,** Overview of the DAbI setup with two LEDs, illuminating an out-of-focus sample from two different angles. The defocus aberrations of the sample under the dual-angle illumination digitally interfere at the overlap region in spectrum, resulting in interference-like fringes. **b,** The physics modeling of interference-like fringes: the amplitude of the intensity spectrum at the overlap region is modulated by a cosine pattern from the defocus aberration difference.

The above derivation and analysis were done assuming coherent light sources. To generalize the mathematical modeling for DAbI and support our experiments, we have extended the theory to partially coherent light sources (Supplementary Note 1.7, 1.8).

## Experimental results

We experimentally validated the performance of DAbI in 2D and 3D autofocusing and digital refocusing and demonstrated DAbI's broad applications in automated optical microscopy. For simplicity in post processing, in all the following experiments, we maintained equal distances between each LED and the optical axis, with illumination angles slightly smaller than the maximum acceptance angle of the imaging system (See Supplementary Note 3).

## Validation of DAbI performance in autofocusing

The capability of autofocusing was assessed by measuring both the accuracy of identifying correct defocus distances and the maximum robust defocus distance achievable. In Fig. 3, various biological and clinical samples were tested (Supplementary Note 10). For the 2D evaluation, samples were intentionally defocused up to 1200 μm on one-side of the focal plane. Nominal defocus distances were obtained using a precision piezo axial stage as ground truth. DAbI consistently maintained high accuracy up to 1800 μm defocus range, approximately 443 times of the DoF (4.06 μm See Methods) under a system setting of 20×/0.40NA objective, a camera pixel pitch of 6.5 μm, LED polar angles $\beta = 23.0°$ and an azimuthal angle $2\theta = 22.5°$. We achieved accuracy within the DoF of the system (Dashed blue line in Fig. 3). When integrated with phase imaging, the error tolerance can be further relaxed (Dashed red line in Fig. 3).

To achieve automation in 3D focusing, the key is to align the central plane of the sample and the focal plane of the objective lens. In Fig. 4, we evaluated 3D autofocusing performance in samples with different types and thicknesses (up to 150-μm). DAbI achieved defocus detection up to 1200 μm range, equivalent to 296 times of the natural DoF (4.06 μm See Methods), under a system setting of 20×/0.40NA objective, a camera pixel pitch of 6.5 μm, LED polar angles $\beta = 23.0°$ and an azimuthal angle $2\theta = 45.0°$. In 3D autofocusing, the accuracy of finding the exact focal plane is more relaxed. For example, the axial-scanning type microscopic imaging scans the 3D sample layer-by-layer, which typically reserves some redundant region at the top and bottom of the sample volume. Here, we empirically set a 10-μm buffer region, approximately 7% – 20% of the sample volume, to accommodate focusing uncertainties in practical 3D imaging workflows.

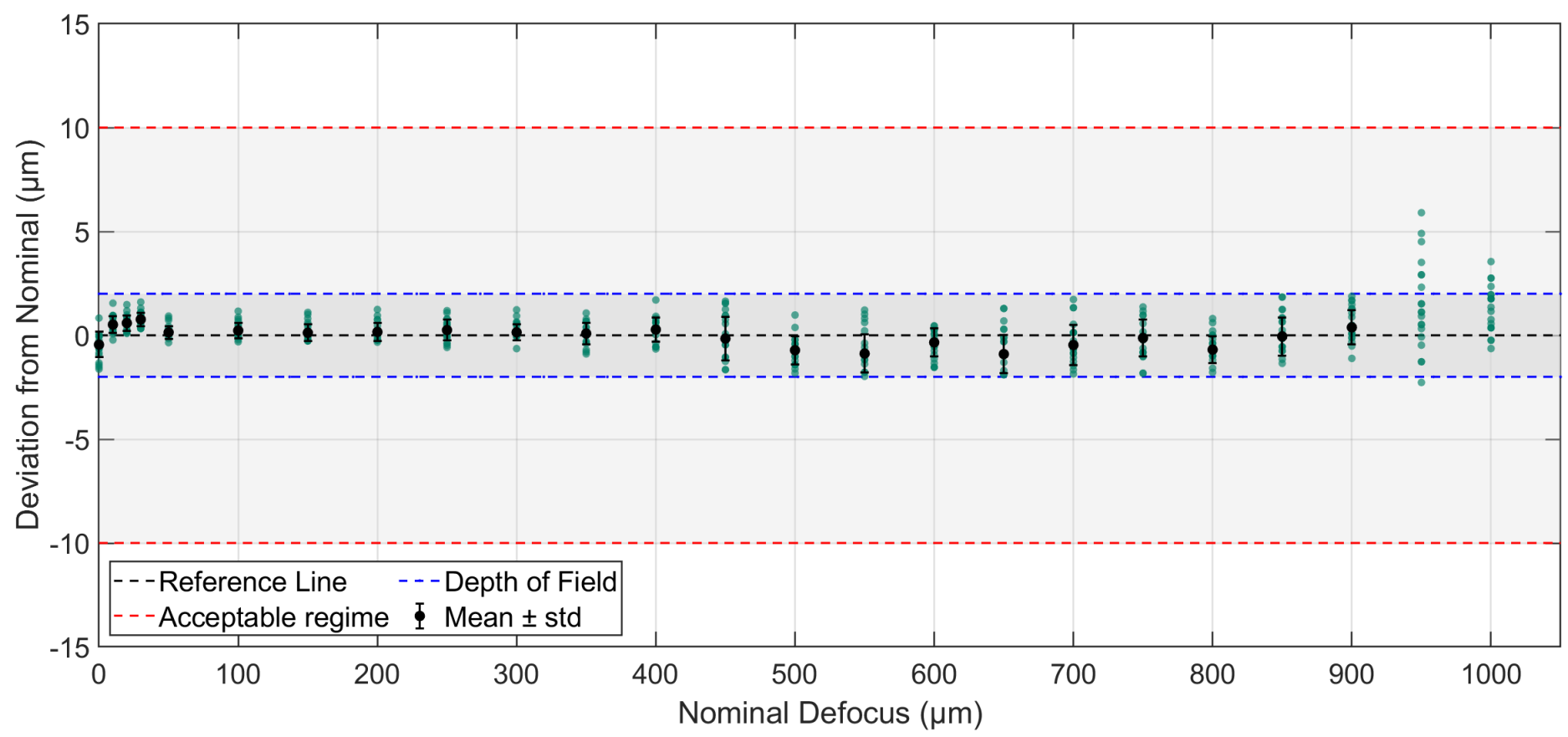

**Fig.3 | Experimental demonstration of focusing capability of DAbI on thin samples.** DAbI focusing ability for thin samples over 20 different samples varied across human tissue sections, cell specimens, and sectioned organoids. The focusing ability can extend to 443-fold of BFM's natural DoF. The blue dashed line indicates the system DoF, corresponding to the desired autofocusing accuracy. The red dashed line denotes the acceptable operating regime when autofocusing is combined with digital refocusing. These experiments are performed on a system setting with 20×/0.40NA objective lens, a camera pixel pitch of 6.5 μm, LED polar angles $\beta = 23.0°$ and an azimuthal angle $2\theta = 22.5°$.

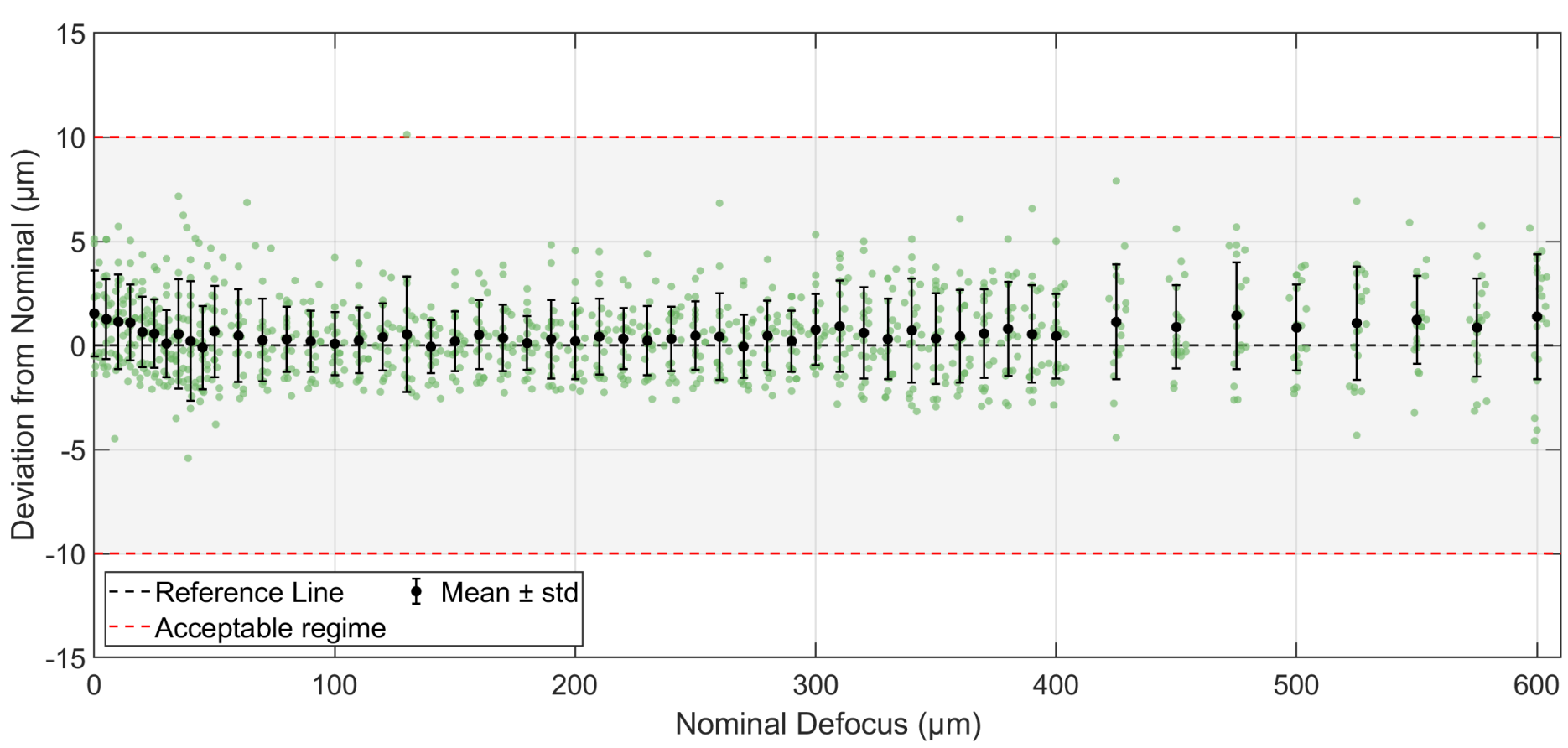

**Fig.4 | Experimental demonstration of focusing capability of DAbI on thick samples.** DAbI focusing ability for thick samples varied from 40 to 150 microns thicknesses with more than 20 sample including mouse tissue sections and organoids. We can extend the autofocusing range to 296-fold of BFM's natural DoF. The red dashed line denotes the acceptable operating regime. These experiments are performed on a system setting with 20×/0.40NA objective lens, a camera pixel pitch of 6.5 μm, LED polar angles $\beta = 23.0°$ and an azimuthal angle $2\theta = 45.0°$.

## DAbI-automated autofocusing for bi-modality fluorescence and phase imaging

Here, we show an application of DAbI with a case study of self-assembling α-synuclein (SAS) transferred HEK cell imaging[44]. Synuclein aggregation is a hallmark feature of Parkinson's disease. Widefield fluorescence and phase imaging provide chemical and structural insights to understand the synuclein expression and its aggregation within the cells.

As shown in Fig. 5, a multi-well slide containing HEK cells with different SAS variants was imaged. Initially, the slide was positioned on the sample holder at well (a). DAbI automatically captured images and calculated a defocus distance $\Delta z$ = -43.5 μm from the fringes. Subsequently, the piezo z-scanner adjusted the sample back to the focal plane, enabling fluorescence and quantitative phase imaging[39] of the correctly focused sample. The x-y scanner then moved to well (b), and DAbI efficiently refocused, compensating for unevenness among different wells. This automated procedure can be systematically applied to other wells and additional multi-well samples. The DAbI-automated bi-modality microscope can characterize synuclein expression and distribution in HEK cell samples to screen promising candidates for modeling and studying Parkinson's disease. Specifically, Fig. 5b shows the green fluorescence protein (b1) aggregated, (b2) expressed but not aggregated, and (b3) unexpressed.

## 3D autofocusing for brightfield imaging, refractive index tomography, and confocal microscopy

For 3D thick samples, we typically have a rough estimate of the sample thickness. For instance, early-stage live mouse embryos generally range from 50 to 100 μm in size, and the thickness of human or animal tissue sections can be approximated based on microtome settings. The key to automated 3D imaging process is the alignment of the central plane of the sample with the focal plane of the objective lens. Since DAbI fringes arise from an average defocus aberration across the 3D volume (Supplementary Note 2), they serve as an excellent indicator for 3D autofocusing.

In Fig. 6, we demonstrate DAbI's 3D autofocusing across three different imaging modalities for live mouse embryo imaging and mouse brain section imaging. The general automated workflow is shown in Fig. 6a. After sample placement, DAbI captures two images and computes the defocus distance from the observed interference-like fringes. The z-scanner then shifts the sample so that its central plane aligns with the focal plane (e.g. $\Delta z$ = +82 μm in Fig. 6a), after which imaging proceeds. This process can be repeated efficiently across multi-well plates or for multiple samples.

Fig. 6b shows label-free quantitative refractive index tomography of a live four-cell stage mouse embryo using analytic Fourier ptychotomography (AFP)[43], with clear visualization of nuclei in the zoomed-in regions (Fig. 6b1, b2). Fig. 6c presents BFM images of another live four-cell mouse embryo following DAbI autofocusing after relocating to a new well.

Fig. 6d shows multi-channel fluorescence images of an 80-μm-thick brain section of an SAS-infected mouse acquired using DAbI-automated confocal microscopy. After actuating the sample to the correct focus, the xz- and yz-projection views of the image volume reveal symmetry about the central plane, confirming its accuracy in determining the defocus detection in 3D specimens.

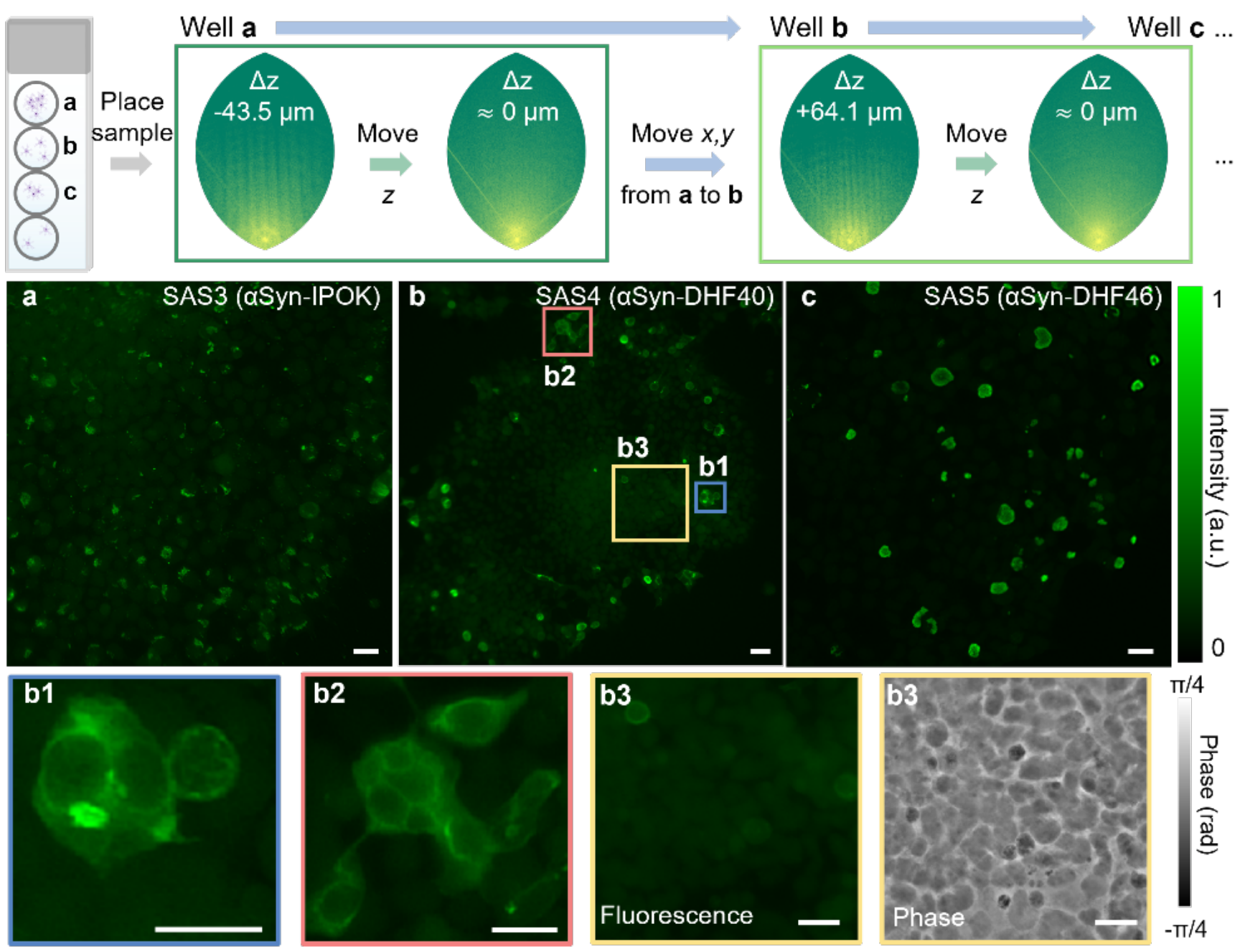

**Fig.5 | DAbI-automated widefield fluorescence microscopy and quantitative phase imaging experiments for self-assembling α-synuclein transfected HEK cell imaging.** Top panel illustrates the general automated imaging flow for four-well (**a-d**) HEK cell sample. **a-c,** Well **a-c** HEK cells transfected with different variants of self-assembling α-synuclein. **b1-2,** Zoom-in image in **b** with protein aggregation (**b1**) and with protein expression but no aggregation (**b2**). **b3,** Unexpressed region in fluorescence imaging in **b**, but with cellular structural information in phase channel. Scale bar: 20 μm

## DAbI performance in digital refocusing by extending depth-of-field for complex-field imaging

Complex-field imaging techniques can digitally refocus images after acquisition. However, they still face significant challenges in correcting large defocus aberrations. DAbI can integrate with these techniques as a defocus prior to digitally resolve defocus aberrations across different levels. Here, we present the integration with the state-of-the-art method, angular ptychographic imaging with closed-form method (APIC)[39], and more demonstration on Fourier ptychographic microscopy is in Supplementary Note 10. APIC reconstructs both the amplitude and phase of the sample from a set of intensity measurements captured under angularly varying illumination, applying a closed-form inversion that jointly retrieves high-resolution complex-field images and system aberrations.

We benchmarked DAbI's capability on a standard USAF1951 phase resolution target (Fig. 7). The achieved resolution was 870 nm with a theoretical value of 814 nm (Method Section). Observing the interference-like fringes arising from defocus aberrations (Fig. 7), we noted that larger defocus values produce denser fringe patterns. The bending of fringes was due to spherical aberrations, but this had minor influence on DAbI's defocus determination capability (Supplementary Note 6.4). With DAbI's defocus detection capability, we further extended APIC's digital refocusing ability by a factor of 4.8, representing a 20-fold increase of effective DoF over conventional brightfield or phase contrast microscopy.

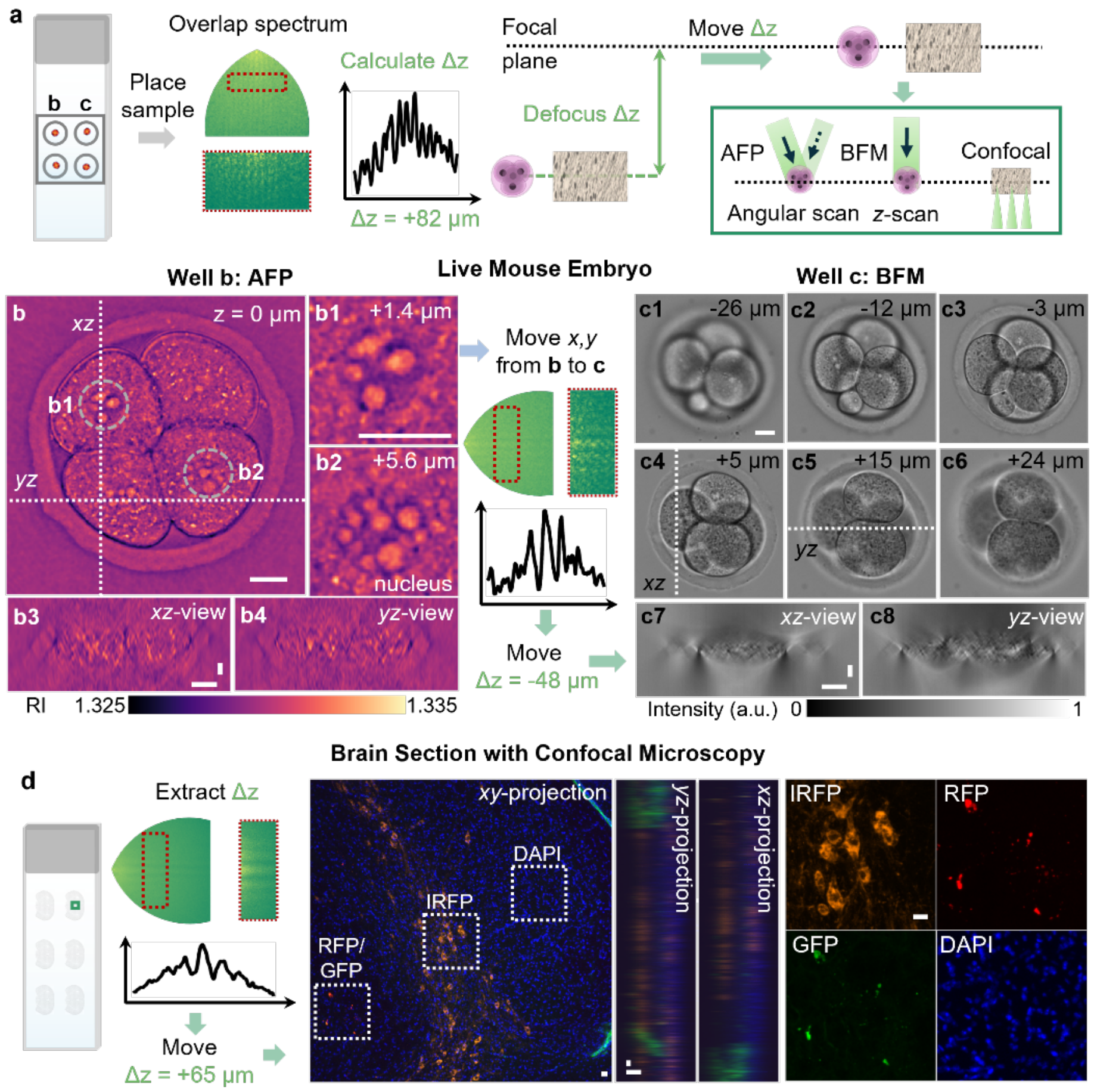

**Fig.6 | Autofocusing experiments for live mouse embryo imaging and 80-µm-thick mouse brain section imaging in refractive index tomography, brightfield transmission microscopy (BFM), and confocal microscopy with DAbI. a,** The general autofocusing for 3D thick samples using DAbI. **b,** Label-free quantitative refractive index (RI) imaging using analytic Fourier ptychotomography (AFP) for a live four-cell stage mouse embryo. **b1,b2,** Zoom-in of embryonic cell nuclei. **b3,b4,** Axial section view of RI reconstructions. **c,** BFM for a four-cell stage mouse embryo imaging with z-scanning. **c1-6,** Images captured at different axial planes. **c7-8,** Axial sectioning of the embryo. **d**, Automated confocal microscopy focusing on an 80-µm-thick mouse brain tissue with four fluorescence channels. IRFP, RFP, GFP are infrared, red, green fluorescence protein respectively. Scale bar: 20 µm

## DAbI-assisted digital refocusing in complex-field imaging

Routine pathology tissue sections are typically on the centimeter scale and prepared in batches, necessitating high-throughput, efficient imaging solutions. We demonstrate the use of DAbI combined with APIC to further enhance digital refocusing and eliminate the need of time-consuming axial scanning during whole slide imaging.

In this context, the pathology samples were positioned, scanned, replaced, and imaged without requiring mechanical refocusing (The top panel of Fig. 8). In Fig. 8a–d, the phase reconstructions of unstained lung and colon cancer tissues reveal that DAbI-assisted APIC significantly improves image clarity and structural detail compared to APIC alone. This improvement is particularly notable in regions exhibiting large defocus variations. In Fig. 8e–g, color images of H&E-stained lung cancer tissues further demonstrate the efficacy of DAbI. The DAbI-assisted images are notably free from blurring and reconstruction artifacts. Moreover, DAbI facilitates compensation for chromatic aberrations, as shown (Supplementary Note 13), yielding high-fidelity color reconstructions across the entire slide. Supplementary Note 11 further demonstrated that DAbI-assisted digital refocusing reconstructions achieve quality comparable to reconstructions from mechanically refocused, in-focus measurements. By correcting aberrations and eliminating the need for physical axial scanning, DAbI-assisted complex-field imaging offers a streamlined, accurate, and scalable solution for digital pathology imaging.

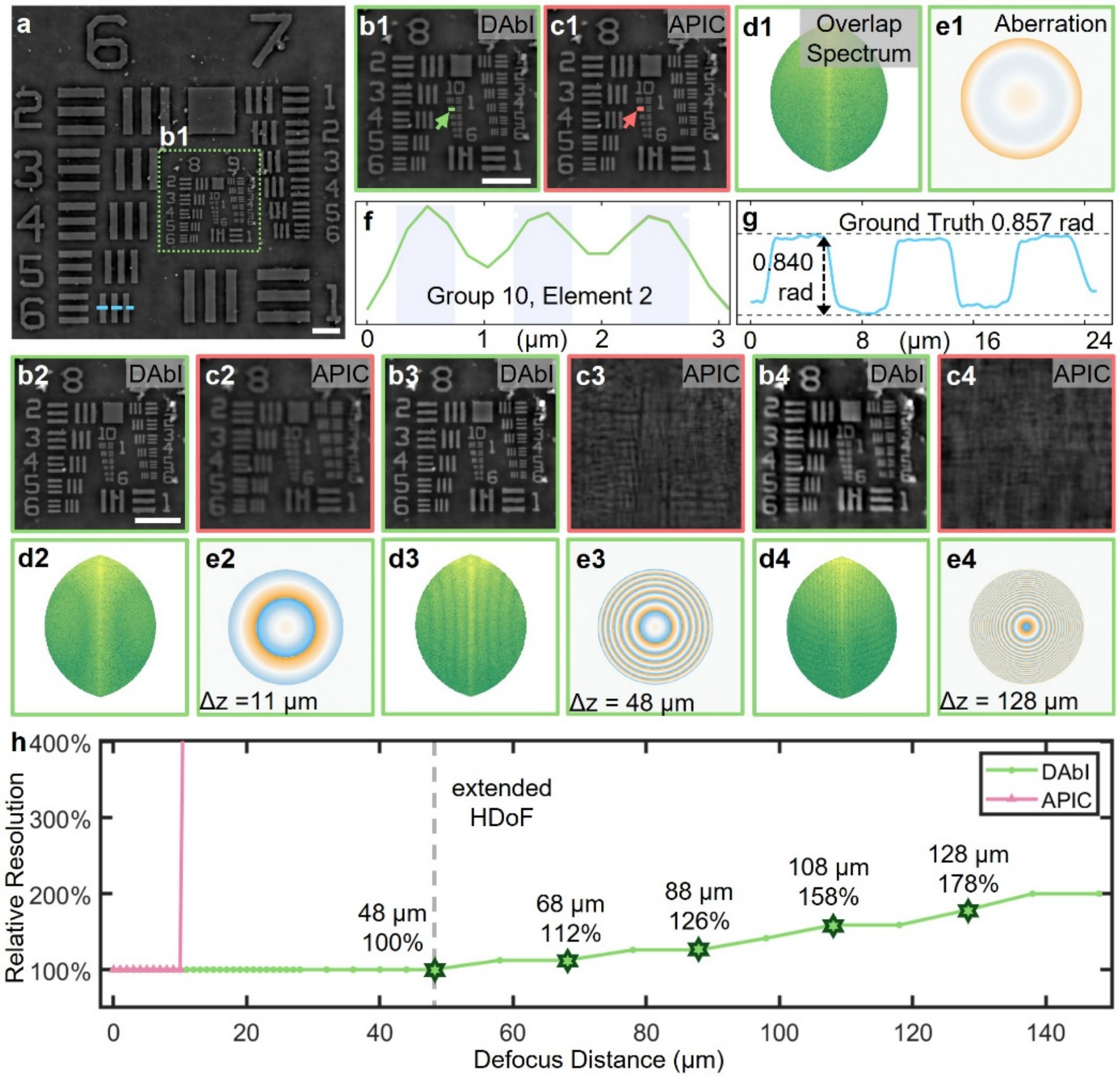

**Fig.8 | Experimental evaluation of digital refocusing with DAbI. a,** USAF1951 phase resolution target in focus with APIC reconstruction. **b1-4,** Zoom-in of reconstructions of USAF1951 phase targe using DAbI-assisted APIC at different defocus distances. **c1-4,** Reconstructions by only using APIC. **d1-4,** The fringe pattern in the amplitude of overlap intensity spectrum at different defocus aberrations. **e1-4,** Pupil phase used in image reconstruction from DAbI-assisted APIC aberration retrieval at corresponding defocus distances. **f,** Line profile of Group 10, Element 2 of the resolution target at the resolution limit of the imaging system. **g,** The phase step of the phase resolution target. Our measured phase value 0.840 rad matched the manufacturer's ground truth 0.857 rad (indicated by gray dashed lines) that obtained through atomic force microscopy **h,** Performance evaluation of digital refocusing with only APIC and DAbI-assisted APIC. HDoF: Half depth-of-field. Scale bar: 20 μm

## Discussion

We presented the discovery and extensive application of the DAbI effect for 2D and 3D autofocusing, as well as for digital refocusing. Compared with existing methods (Supplementary Note 9), DAbI stands out with its long-range capability, system simplicity, low light toxicity, high efficiency, and generalizability. With DAbI employed in our specific systems, we achieved ~1800 µm autofocusing range (443-fold DoF) in 2D and ~1200 µm (~296-fold DoF) in 3D. When integrated with complex-field imaging, a 20-fold extension of natural DoF was achieved through digital refocusing. The exceptional performance was realized through a simple two-LED configuration, which is readily adaptable to both customized and commercial microscopy systems. We demonstrated DAbI's wide applications across brightfield, color, refractive index, complex-field, confocal, and widefield fluorescence microscopes. Moreover, DAbI possesses great potential in facilitating or automating holographic microscopy[45,46], single molecule localization microscopy[47], and structured illumination microscopy[48,49].

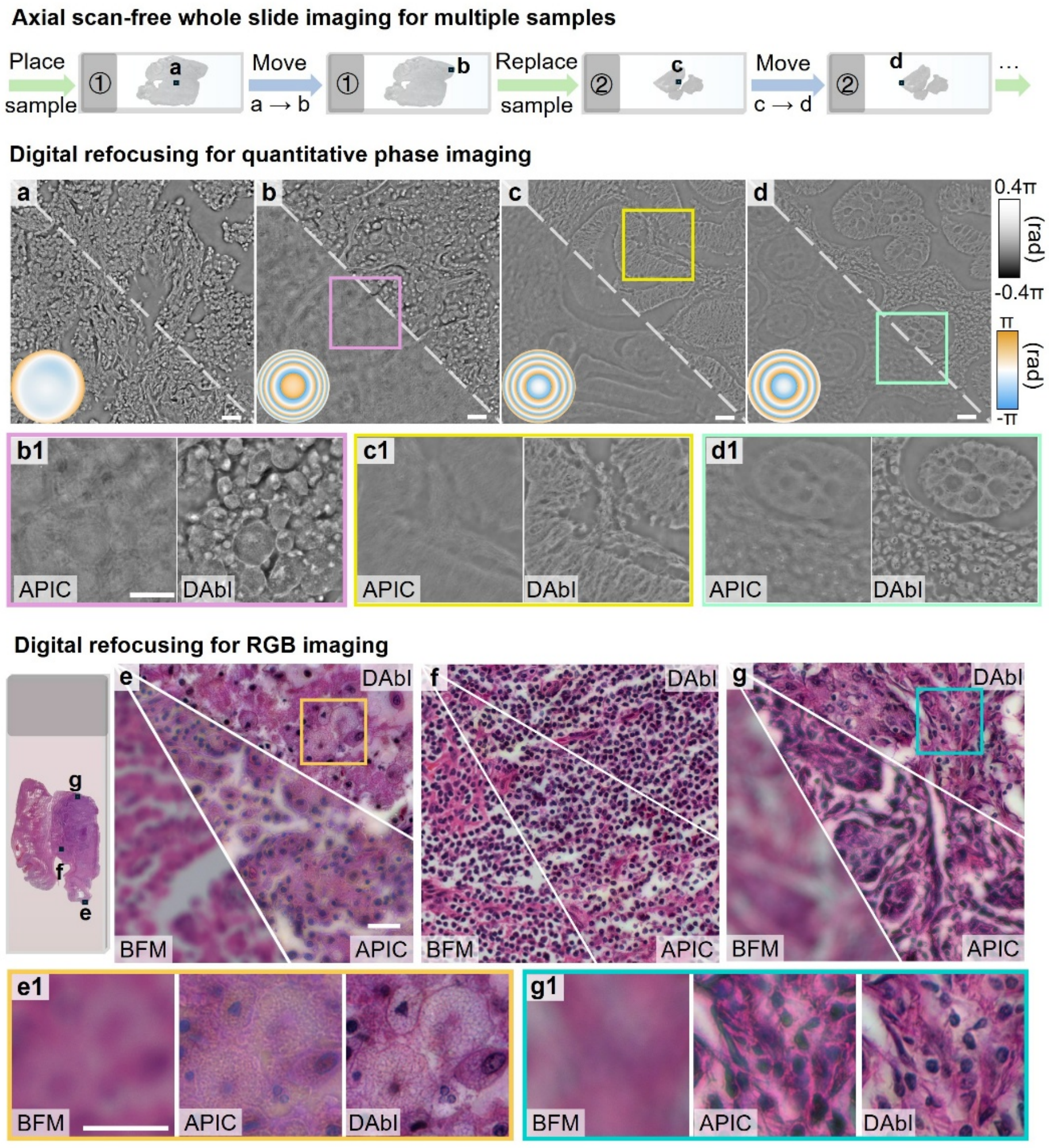

**Fig.7 | DAbI digital refocusing experiments in complex-field imaging for digital pathology whole slide imaging.** Top panel illustrates the general axial scan-free imaging workflow for multiple whole slide samples. **a-d,** Phase images of unstained human lung cancer and colon cancer tissues from different locations of two slides with corresponding defocus aberrations. **b1,c1,d1,** Zoom-in images in (**b-d**) with a comparison without/with DAbI assistance. **e-g,** Absorption-based imaging for H&E-stained non-small-cell lung cancer tissue sections with a comparison among conventional brightfield transmission microscopy (BFM), APIC, and DAbI-assisted APIC. **e1,g1,** Zoom-in images in **e** and **g**. Scale bar: 30 µm

In addition to its strong capability and broad applications, DAbI excels in efficiency, both in acquisition and computational feedback. In our demonstrated experiments, only two images were required to implement DAbI, captured under 40 ms with an illumination power density of ~2 µW/mm$^2$ in the sample plane. The low photon budget requirement of DAbI makes it well-suited for live imaging or light-sensitive samples. On the computational side, the feedback time was approximately 0.34 s on a CPU (Intel i7-14700KF) and 0.11 s on a GPU (Nvidia GeForce RTX4090) for a patch size of 1536 pixels (Detailed computational time analysis in Supplementary Note 12). The rapid acquisition and feedback support its use in high-throughput imaging. From the algorithm perspective, the major computational load comes from fast Fourier transform with a complexity of $\mathcal{O}(N \log N)$, where $N$ is the total image pixel counts. Deep learning methods could potentially further expedite fringe valley detection in DAbI.

Due to its physics-based nature, DAbI performs reliably and consistently across diverse biomedical sample types, including 2D, 3D, stained, and label-free specimens. However, we note that DAbI relies on the visibility of fringes, which is affected by thickness and scattering properties of the sample (Supplementary Note 2, 7). If the sample is highly scattered or very thick (e.g. >200 μm), the contrast of the fringes will be very low, thus, imposing a challenge to finding the defocus distance. This condition can be mitigated through optical clearing techniques[50,51], to enable DAbI's application to thicker samples. Additionally, longer illumination wavelengths can reduce the impact of multiple scattering, further expanding DAbI's operational regime.

The autofocusing range of DAbI has an upper limit given by the sampling rate in frequency domain, which is mainly determined by the field-of-view size of the imaging system (Supplementary Note 6.2). However, increasing the field-of-view may result in heavier computational load. Thus, a balance between autofocusing range and computational efficiency should be considered based on the downstream application requirements.

In this work, DAbI is primarily applied to biological and clinical samples whose amplitude spectra do not exhibit specific structures. In Supplementary Note 8, we further experimentally validate the applicability of DAbI to structured samples, such as Siemens stars or phase gratings.

Beyond autofocusing, DAbI holds promise in sample thickness estimation and corrections for other aberrations. In 3D autofocusing, the visibility range of the fringes in DAbI is associated with the sample thickness (Supplementary Note 7). By evaluating the contrast of the fringes, the sample thickness can be potentially estimated. Furthermore, when a dominant aberration (e.g., spherical or coma) exists in the imaging system, the principles of DAbI may be adapted to retrieve and correct such aberrations.

Although all demonstrations of DAbI in this work were performed on transmission optical microscopes, the method is equally compatible with reflection-type systems. This extends its utility beyond biomedical imaging to automated industrial inspection. We outline a potential configuration for implementing DAbI on reflection-mode microscopes in Supplementary Note 14. To conclude, the observation and utilization of DAbI bring new insight into optical imaging and microscope automation. With prominent advantages in simplicity, robustness, high-efficiency, generalizability, and superior performance, DAbI shows great potential in enabling fully automated optical microscopy. When integrated with robotic sample handling[52,53] and deep learning-based sample region detection[54,55], DAbI can pave the way for a fully human-free automated imaging workflow in the future.

## Methods

### Imaging systems

Three imaging platforms were employed in this study. The primary imaging system integrated a customized, programmable LED array (Red: XQARED-00-0000-000000Y01, Green: XQABLU-00-0000-000000T02, Blue: XQAGRN-00-0000-000000Y01) with a commercial inverted microscope (Olympus IX71), equipped with an objective lens (Olympus Plan N 20×/0.40 NA) and an sCMOS camera (Excelitas Inc. pco.edge 5.5 CLHS, 6.5 μm pixel pitch). A fluorescence imaging path was incorporated to enable widefield fluorescence acquisition with a 470 nm excitation LED (Thorlabs, M470L5), a GFP excitation filter (Thorlabs MF469-35), and a GFP emission filter (Thorlabs MF525-39). In this work, this system was used for autofocusing performance evaluation (Figs. 3,4), super extended DoF quantification (Fig. 7), widefield fluorescence microscopy (Fig. 5), quantitative complex-field imaging (Fig. 5,8), refractive index tomography (Fig. 6), and BFM (Fig. 6).

DAbI can also be enabled by using the LED array with other commercial imaging systems. The microscope system for H&E-stained pathology imaging was built upon a commercial inverted microscope (Olympus IX51) with a motorized translation stage (Thorlabs MLS203-1) for xy-scanning, an objective lens (Olympus Plan N 20×/0.40 NA), and a CMOS camera (Allied Vision Prosilica GT6400). This system was used for absorption-based color imaging for H&E-stained non-small-cell lung cancer tissue sections (Fig. 8). The confocal system was built on a commercial confocal microscope (Andor, Dragonfly 200) with an objective lens (HC PL FLUOTAR 10×/0,30 NA) and a camera (Andor ZL41 Cell sCMOS). Details about system calibration are discussed in Supplementary Note 5.

## Definition of depth-of-field

The depth-of-field (DoF) of the conventional brightfield transmission microscope or phase contrast microscope can be empirically evaluated by[56],

$$\mathrm{DoF} = \frac{\lambda n_{\mathrm{media}}}{\mathrm{NA}^2} + \frac{n_{\mathrm{media}} e}{M \cdot \mathrm{NA}}$$

where $\lambda$ is the illumination wavelength (peak wavelength of the narrow-bandwidth LED); $n_{\mathrm{media}}$ is the refractive index of the background medium; NA stands for numerical aperture; $e$ denotes the camera pixel size; $M$ is the magnification of the imaging system. Given parameters for the experiments in Figs. 3,4, $\mathrm{DoF} = \frac{520\ \mathrm{nm} \times 1}{0.4^2} + \frac{1 \times 6.5\ \mu\mathrm{m}}{20 \times 0.40} = 4.06\ \mu\mathrm{m}$, with a half DoF of 2.03 μm. In Fig. 7, $\mathrm{DoF} = \frac{635\ \mathrm{nm} \times 1}{0.4^2} + \frac{1 \times 6.5\ \mu\mathrm{m}}{20 \times 0.40} = 4.78\ \mu\mathrm{m}$, with a half DoF of 2.39 μm.

## Resolution in complex-field imaging

For angular scanning type complex-field imaging, such as FPM or APIC, the resolution of the imaging system is determined by the NA of objective lens and the maximum illumination angle[32]. In Fig. 7, the theoretical resolution is calculated by $\frac{\lambda}{\mathrm{NA} + \mathrm{NA}_{\mathrm{illu}}} = \frac{635\ \mathrm{nm}}{0.4 + 0.38} = 814\ \mathrm{nm}$.

## Sample preparation

**USAF1951 resolution target.** The USAF1951 phase resolution target (Benchmark Technologies Inc.) has a height difference of 166.6 nm, measured by an atomic force microscope from the manufacture. The phase target has a refractive index of 1.52 for the material, resulting in a phase jump of 0.857 rad. The full-pitch resolution of the phase target is given by $\frac{1000\ \mu\mathrm{m}}{2^{\mathrm{Group} + (\mathrm{Element} - 1)/6}}$. Given Group 10 Element 2 was resolved in Fig. 7, the resulting resolution is 870 nm.

**HEK cells and proteins.** Self-assembling α-synuclein (SAS) plasmids were transfected into HEK cells using Lipofectamine 3000 (ThermoFisher, L3000008) following the manufacture's instruction. Briefly, 100K HEK cells were plated in 24-well plates. The next day, 0.4 μg SAS plasmids and 0.1 μg rtTA plasmids were transfected into each well. Cells were incubated with lipofectamine overnight. The medium was switched to 5% FBS in DMEM/F12 (ThermoFisher, Cat #21331020) the next morning, and 10 μg/ml of doxycycline (Sigma, Cat # D9891-10G) was added into the medium. After 3 days of doxycycline treatment, cells were fixed and stained with V5 tag for detecting the SAS (Abcam, Cat # ab309485).

**Human cancer histology samples.** The H&E-stained and unstained human lung cancer tissues sections were prepared from formalin-fixed paraffin-embedded (FFPE) tissue blocks. The tissue blocks were first cut into 5-μm tissue sections, unfolding in a water tank. Then, the samples went through deparaffinized (xylene) and dehydrated (95% and 100% ethonal) processes, while the H&E-stained samples had an extra staining step for hemotoxylin and eosin stains, respectively. The colon cancer sample was purchased from Acepix Bioscinces Inc. with the same preparation for unstained sample from FFPE tissue blocks.

**Mouse embryos.** All animal works were approved by the IACUC. 6-week ages of B6D2F1/J female mice (Jackson Laboratory) were used for this study. Female mice were super-ovulated by post-pregnant mare's serum gonadotrophin (PMSG) and human chorionic gonadotrophin (hCG) injection. Zygotes were collected in M2 medium 22 hours after hCG injection and subsequently cultured in Advanced KSOM medium at 37 °C with 5% CO2 in the air. Four-cell live embryos were used for this study. The live embryos were transferred in 7 μL of KSOM medium ($n \approx 1.33$) for imaging. The embryos were placed in sperate wells of adhesive silicone isolators (9 mm in diameter and 0.8 mm in depth), covered with a coverslip and sealed with nail polish.

**Mouse Tissue harvesting and processing.** Mice were transcranial perfused with 30 mL of ice-cold heparinized 1× PBS, and brains and other organs were dissected. One hemisphere of the brain was used for RNAseq. For analysis of fluorescent protein expression, one hemisphere of the brain and other organs was submerged in ice-cold 4% PFA formulated in 1× PBS and fixed for 24-48 hr at 4 °C. Brains and other organs were subsequently cryoprotected at 4°C in a solution containing 30% (w/v) sucrose for 72 hr and then were flash-frozen in O.C.T. Compound (Scigen, Cat # 4586) using a dry ice-ethanol bath and kept at -80 °C until sectioning. Tissues were sliced at different thicknesses based on experimental needs using a cryostat (Leica Biosystems, CM1950) and kept in PBS with sodium aside.

**Human brain organoids.** Human embryonic stem cells (hESC) were plated on 6-well plate coated with Vitronectin (Gibco, Cat # A14700) at 1:100 dilution in DPBS (Gibco, Cat #14190250) and maintained in E8 medium (Thermo Fisher, Cat #A1517001). The E8 medium was changed every day, and the cells were passaged every 3-5 days at 70-85% confluence. The cells were passaged using EDTA dissociation. For inducing differentiation, hESCs were dissociated at 70%-80% confluence using EDTA dissociation buffer and replated as a single cell suspension on Matrigel-coated 24-well plates at a density of 100K cells/cm2 in E6 medium (Thermo Fisher, Cat #A1516401) containing 10 μM ROCK-inhibitor (Y-27632; R&D, Cat #1254). The cells were kept in E6 medium + Y-27632 overnight. The next day, the medium was switched to a neuroectoderm-inducing medium. The cells were kept in E6 medium containing 10 μM SB431542 (R&D, Cat #1614), 100nM LDN139189 (R&D, Cat #6053), and 500nM XAV (R&D, 3748) for the first 4 days and then switched to E6 medium containing 10 μM SB and 100nM LDN for 6 days. Then the cells were switched to progenitor expanding medium, containing neurobasal (NB) medium supplemented with l-glutamine (Gibco, Cat #25030-164), N2 (Stem Cell Technologies, 07156), B27 (Life Technologies, Cat #17504044) and NEAA (Sigma, Cat #M7145) for 10 days. At D20, the neuron progenitors were dissociated into single cells and replated into a V-bottom ultra-low attachment 96-well plate at 100K per well to form neuron spheres. After 4 days, the spheres were transformed to low attachment 10 cm dishes and placed on a shaker for long-term culture and maturation in cortical neuron medium containing NB/N2/B27/Glu/NEAA, ascorbic acid (200μM, Sigma, 4034-100g), dbcAMP (500 μM, Sigma, Cat #D0627), BDNF (20ng/mL, R&D, Cat #248-BDB) and GDNF (20ng/mL, Peprotech, Cat #450-10).

**Immunostaining.** Brain or human organoids sections were blocked with 5% normal donkey serum in 0.1 % PBST for 90 mins, then incubated with primary antibody solution prepared in 0.1 % PBST supplemented with 5% normal donkey serum overnight on a shaker at 4°C. After washing in 0.1 % PBST (3 × 30 min), the sections were incubated with secondary antibodies overnight on a shaker at 4°C and then washed 3 x 30 min in 0.1 % PBST. The tissues were mounted on glass slides with Prolong Diamond Antifade mounting media (ThermoFisher Scientific, Cat # P36970).

## Data availability

The data that support this study are publicly available at https://osf.io/dvztc/

## Code availability

The code for DAbI is available at https://github.com/hwzhou2020/DAbI

## Acknowledgements

H.Z., S.Z., Z.D., O.Z. and C.Y. are supported by Rothenberg Innovation Initiative ($RI^2$) (Award number A4188-Yang-3-A1) and the Heritage Medical Research Institute (HMRI) (Award number HMRI-15-09-01). Y.F. is supported by Caltech Chen Postdoc Innovator Grant (Award Number ENDOW.CHEN-1.CPIACY25). H.Z. thanks for the support of Caltech Schmidt Graduate Research Fellowship. We thank Dr. Shoma Nakagawa from Caltech for preparing the live mouse embryo samples. We thank Prof. Richard J. Cote and Prof. Mark Watson from the School of Medicine at Washington University in Saint Louis for providing human lung cancer specimens.

## Author contributions

H.Z. and S.Z. conceived the idea. H.Z. and S.Z. conducted theoretical analysis, simulations and experiments. Y.F. prepared biological samples and helped with experimental designs and implementations and V.G. supervised this part. Z.D. helped with theory development and experimental designs. O.Z. helped with the fluorescence experiments. C.Y. guided the project design and supervised this project. All authors contributed to the preparation of the manuscript.

## Competing interests

The authors declare the following competing interests: On July 2, 2025, California Institute of Technology filed a provisional patent application (CIT-9339-P) for the DAbI method, which covered the concept, implementation, and applications of the DAbI method described here.

## Supplementary information

A supplementary file (Notes 1-15) and a supplementary video are attached to this work.